\documentclass[journal]{IEEEtran}
\usepackage{amsthm}
\usepackage{amsmath}
\usepackage{wasysym}
\usepackage{rotating}
\usepackage{wasysym}
\usepackage{url}
\usepackage{hyperref}
\newtheorem{mydef}{Definition}
\newtheorem{myrem}{Remark}

\begin{document}
\title{Inference rules for RDF(S) and OWL in N3Logic}

\author{\IEEEauthorblockN{Dominik Tomaszuk}
\IEEEauthorblockA{Institute of Computer Science\\
University of Bialystok\\
Poland\\
Email: \href{mailto:dtomaszuk@ii.uwb.edu.pl}{\nolinkurl{dtomaszuk@ii.uwb.edu.pl}}}}

\maketitle

\begin{abstract}
This paper presents inference rules for Resource Description Framework (RDF), RDF Schema (RDFS) and Web Ontology 
Language (OWL). Our formalization is based on Notation 3 Logic, which extended RDF by logical symbols and 
created Semantic Web logic for deductive RDF graph stores. We also propose OWL-P that is a lightweight formalism of
OWL and supports soft inferences by omitting complex language constructs.
\end{abstract}

\IEEEpeerreviewmaketitle

\section{Introduction}
Resource Description Framework (RDF) is a general method for conceptual description or modeling of information that is implemented 
in web resources. RDF Schema (RDFS) extends RDF to classes providing basic elements for the description of vocabularies. OWL adds 
more vocabulary for describing properties and classes i.e. relations between classes, cardinality, and richer typing of properties.
Unfortunately, OWL has high worst-case complexity results for key inference problems. To overcome this problem we propose a
lightweight OWL profile called OWL-P.

A rule is perhaps one of the most understandable notion in computer science. It consists of the condition and the conclusion. 
If some condition that is checkable in some dataset holds, then the conclusion is processed. In the same way RDF(S) and OWL
entailments work.

The paper is constructed according to sections. Section~\ref{sec:preliminaries} presents RDF and Notation 3 Logic concepts. 
In Section~\ref{sec:rules} we present inference rules for RDF. RDFS and OWL in N3Logic. Section~\ref{sec:relatedwork} is devoted 
to related work. The paper ends with conclusions.

\section{Preliminaries}
\label{sec:preliminaries}
The RDF data model rests on the concept of creating web-resource statements in the form of subject-predicate-object expressions, 
which in the RDF terminology, are referred to as \emph{triples} (or \emph{statements}).

An RDF triple comprises a subject, a predicate, and an object. In \cite{Wood2014}, the meaning of subject,
predicate and object is explained. The \emph{subject} denotes a resource, the \emph{object} fills the value of the relation, 
the \emph{predicate} refers to the resource's characteristics or aspects and expresses a subject -- object relationship. 
The predicate denotes a binary relation, also known as a property.

Following \cite{Wood2014}, we provide definitions of RDF triples below.

\begin{mydef}[RDF triple]
\label{def:rdftriple}
Assume that $\mathcal{I}$ is the set of all Internationalized Resource Identifier (IRI) 
references, $\mathcal{B}$ (an infinite) set of blank nodes, $\mathcal{L}$ the set of literals.
An \emph{RDF triple} $t$ is defined as a triple $t = \langle s, p, o\rangle$ where $s \in \mathcal{I} \cup \mathcal{B}$ is 
called the \emph{subject}, $p \in \mathcal{I}$ is called the \emph{predicate} and $o \in \mathcal{I} \cup \mathcal{B} \cup \mathcal{L}$ 
is called the \emph{object}.
\end{mydef}

The elemental constituents of the RDF data model are RDF terms that can be used in reference to resources: 
anything with identity. The set of RDF terms is divided into three disjoint subsets: IRIs, literals, and blank nodes.

\begin{mydef}[IRIs]
\label{def:rdfterms}
\emph{IRIs} serve as global identifiers that can be used to identify any resource.
\end{mydef}

\begin{mydef}[Literals]
\label{def:rdfliterals}
\emph{Literals} are a set of lexical values. It can be a set of plain strings, such as \verb|"Apple"|, optionally 
with an associated language tag, such as \verb|"Apple"@en|.
\end{mydef}

\begin{myrem}
In RDF 1.1 literals comprise a lexical string and a datatype, such as \verb|"1"^^http://www.w3.org/2001/XMLSchema#int|. 
\end{myrem}

\begin{myrem}
In literals datatypes are identified by IRIs, where RDF borrows many of the datatypes defined in XML Schema 1.1 \cite{Sperberg-McQueen2012}.
\end{myrem}

\begin{mydef}[Blank nodes]
\label{def:bnodes}
\emph{Blank nodes} are defined as existential variables used to denote the existence of some resource for which an IRI or 
literal is not given.
\end{mydef}

\begin{myrem}
Blank nodes are inconstant or stable identifiers and are in all cases locally scoped to the RDF store or the RDF file.
\end{myrem}

A collection of RDF triples intrinsically represents a labeled directed multigraph. The nodes are the subjects and objects of 
their triples. RDF is often referred to as being \emph{graph structured data} where each $\langle s, p, o\rangle$ triple can be 
interpreted as an edge $s \xrightarrow{p} o$.

\begin{mydef}[RDF graph]
\label{def:rdfgraph}
Let $\mathcal{O} = \mathcal{I} \cup \mathcal{B} \cup \mathcal{L}$ and
$\mathcal{S} = \mathcal{I} \cup \mathcal{B}$, then $G \subset \mathcal{S} \times \mathcal{I} \times \mathcal{O}$ 
is a finite subset of \emph{RDF triples}, which is called \emph{RDF graph}.
\end{mydef}

On the other hand, in the Semantic Web environment there is a Notation3 format, which offers a new human-readable serialization of 
RDF model but it also extended RDF by logical symbols and created a new Semantic Web logic called Notation3 Logic (N3Logic). 
Following \cite{Arndt2015}, we provide definitions of N3Logic below.

\begin{mydef}[N3Logic alphabet]
A \emph{N3Logic alphabet} $A_{N3}$ consists of the following disjoint classes of symbols:
\begin{enumerate}
 \item a set $\mathcal{I}$ of IRI symbols beginning with \verb|<| and ending with \verb|>|,
 \item a set $\mathcal{L}$ of literals beginning and ending with \verb|"|,
 \item a set $\mathcal{V}$ of variables, $\mathcal{V} = \mathcal{B} \cup V_U$, where $B$ is a set of existential variables 
 (blank nodes in RDF-sense) start with \verb|_:| and $V_U$ is a set of universal variables start with \verb|?|,
 \item brackets \verb|{|, \verb|}|,
 \item a logical implication \verb|=>|,
 \item a period \verb|.|,
 \item a period \verb|@false|.
\end{enumerate}
\end{mydef}

\begin{myrem}
Notation3 allows to abbreviate IRIs by using prefixes. Instead of writing \verb|<http://example.com>|, we can write \verb|ex:|.
\end{myrem}

\begin{myrem}
Each IRI, variable and literal is an expression.
\end{myrem}

\begin{myrem}
\texttt{\{$f$\}} is an expression called formula.
\end{myrem}

\begin{myrem}
\texttt{$e_1$ => $e_2$} is a formula called implication.
\end{myrem}

In Notation3 literals, IRIs, variables or even formula expressions can be subjects, objects or predicates.

\section{Inference rules}
\label{sec:rules}
In this section, we introduce inference rules for RDF, RDFS and OWL. Inference rules connected with RDF(S) and OWL are basis of the 
deductive RDF graph stores.

\begin{mydef}[Deductive RDF graph store]
A \emph{deductive RDF graph store} is an entity which remembers RDF triples and can generate new ones under certain conditions 
through deduction or inference. It can answer queries about the combined given and inferred triples.
\end{mydef}

\subsection{RDF and RDFS}
In Table~\ref{tab:rulesrdf} we present patterns which hold by RDF and RDFS entailments. All rules are tested in reasoning engines such as 
FuXi\footnote{\url{https://github.com/RDFLib/FuXi}} and cwm\footnote{\url{http://www.w3.org/2000/10/swap/doc/cwm.html}}.

\begin{table*}[]
\caption{Inference rules for RDF and RDFS}
\label{tab:rulesrdf}
\centering
\begin{tabular}{|lll|}
\hline
Conditions &  & Conclusions \\
\hline
\verb|{?S ?P ?O}| & \verb|=>| & \verb|{?P rdf:type rdf:Property}.| \\
\verb|{?P rdfs:domain ?C. ?S ?P ?O}| & \verb|=>| & \verb|{?S rdf:type ?C}.| \\ 
\verb|{?P rdfs:range ?C. ?S ?P ?O}| & \verb|=>| & \verb|{?O rdf:type ?C}.| \\ 
\verb|{?S ?P ?O}| & \verb|=>| & \verb|{?S rdf:type rdfs:Resource}.| \\ 
\verb|{?S ?P ?O}| & \verb|=>| & \verb|{?O rdf:type rdfs:Resource}.| \\ 
\verb|{?Q rdfs:subPropertyOf ?R. ?P rdfs:subPropertyOf ?Q}| & \verb|=>| & \verb|{?P rdfs:subPropertyOf ?R}.| \\
\verb|{?Q rdf:type rdf:Property}| & \verb|=>| & \verb|{?Q rdfs:subPropertyOf ?Q}.| \\ 
\verb|{?P rdfs:subPropertyOf ?R. ?S ?P ?O}| & \verb|=>| & \verb|{?S ?R ?O}.| \\ 
\verb|{?C rdf:type rdfs:Class}| & \verb|=>| & \verb|{?C rdfs:subClassOf rdfs:Resource}.| \\ 
\verb|{?A rdfs:subClassOf ?B. ?S rdf:type ?A}| & \verb|=>| & \verb|{?S rdf:type ?B}.| \\ 
\verb|{?Q rdf:type rdfs:Class}| & \verb|=>| & \verb|{?Q rdfs:subClassOf ?Q}.| \\ 
\verb|{?B rdfs:subClassOf ?C. ?A rdfs:subClassOf ?B}| & \verb|=>| & \verb|{?A rdfs:subClassOf ?C}.| \\ 
\verb|{?X rdf:type rdfs:ContainerMembershipProperty}| & \verb|=>| & \verb|{?X rdfs:subPropertyOf rdfs:member}.| \\ 
\verb|{?X rdf:type rdfs:Datatype}| & \verb|=>| & \verb|{?X rdfs:subClassOf rdfs:Literal}.| \\ 
\hline
\end{tabular}
\end{table*}

\subsection{OWL}

In Table~\ref{tab:surowl} we analyze existing proposals for different OWL2 profiles: RDFS++ \cite{Rdfspp2016}, 
L2 \cite{Fischer2010}, RDF 3.0/OWLPrime \cite{Hendler2010}, OWLSIF/pD* \cite{Terhorst2005}, OWL-LD \cite{Glimm2011} and 
OWL-RL \cite{Motik2012}. We check which terms are most commonly used and propose a new version of OWL 2 called OWL-P. 
We also considered time complexity for detecting a required rule application and frequently used vocabulary terms in 
our corpus. The snapshot (Table \ref{tab:fowl}) is built by \cite{Isele2010} and use seeds from \cite{Schmachtenberg2014}. 

This profile of OWL2 is simpler that OWL-RL. It drops support for restriction and cardinality classes, class 
relationships and list-based axioms. In the Table~\ref{tab:rulesowl1}, Table~\ref{tab:rulesowl2} and we present 
inference rules of OWL-P.

\begin{table*}[]
\caption{Comparison of OWL profiles}
\label{tab:surowl}
\centering
\begin{tabular}{|l|c|c|c|c|c|c|c|}
\hline
                          & OWL-P & RDFS++ & L2  & RDFS 3.0              & OWLSIF         & OWL-LD & OWL-RL \\
                          &       &        &     & OWLPrime              & pD*            &        &        \\
\hline
\verb|owl:AllDifferent|              & \XBox         & \XBox          & \XBox       & \CheckedBox                    & \XBox                   & \XBox          & \CheckedBox    \\
\verb|owl:AllDisjointClasses|        & \XBox         & \XBox          & \XBox       & \XBox                          & \XBox                   & \XBox          & \CheckedBox    \\
\verb|owl:AllDisjointProperties|     & \XBox         & \XBox          & \XBox       & \XBox                          & \XBox                   & \XBox          & \CheckedBox    \\
\verb|owl:allValuesFrom|             & \XBox         & \XBox          & \XBox       & \XBox                          & \CheckedBox             & \XBox          & \CheckedBox    \\
\verb|owl:assertionProperty|         & \XBox         & \XBox          & \XBox       & \XBox                          & \XBox                   & \XBox          & \CheckedBox    \\
\verb|owl:AsymmetricProperty|        & \CheckedBox   & \XBox          & \XBox       & \XBox                          & \XBox                   & \CheckedBox    & \CheckedBox    \\
\verb|owl:cardinality|               & \XBox         & \XBox          & \XBox       & \XBox                          & \XBox                   & \XBox          & \CheckedBox    \\
\verb|owl:complementOf|              & \XBox         & \XBox          & \XBox       & \XBox                          & \XBox                   & \CheckedBox    & \CheckedBox    \\
\verb|owl:DatatypeProperty|          & \CheckedBox   & \XBox          & \XBox       & \CheckedBox                    & \XBox                   & \CheckedBox    & \CheckedBox    \\
\verb|owl:differentFrom|             & \CheckedBox   & \XBox          & \XBox       & \CheckedBox                    & \CheckedBox             & \CheckedBox    & \CheckedBox    \\
\verb|owl:disjointUnionof|           & \XBox         & \XBox          & \XBox       & \XBox                          & \XBox                   & \XBox          & \CheckedBox    \\
\verb|owl:disjointWith|              & \CheckedBox   & \XBox          & \XBox       & \CheckedBox                    & \CheckedBox             & \CheckedBox    & \CheckedBox    \\
\verb|owl:equivalentClass|           & \CheckedBox   & \XBox          & \CheckedBox & \CheckedBox                    & \CheckedBox             & \CheckedBox    & \CheckedBox    \\
\verb|owl:equivalentProperty|        & \CheckedBox   & \XBox          & \CheckedBox & \CheckedBox                    & \CheckedBox             & \CheckedBox    & \CheckedBox    \\
\verb|owl:FunctionalProperty|        & \CheckedBox   & \XBox          & \XBox       & \CheckedBox                    & \CheckedBox             & \CheckedBox    & \CheckedBox    \\
\verb|owl:hasKey|                    & \XBox         & \XBox          & \XBox       & \XBox                          & \XBox                   & \XBox          & \CheckedBox    \\
\verb|owl:hasSelf|                   & \XBox         & \XBox          & \XBox       & \XBox                          & \XBox                   & \XBox          & \CheckedBox    \\
\verb|owl:hasValue|                  & \XBox         & \XBox          & \XBox       & \XBox                          & \CheckedBox             & \XBox          & \CheckedBox    \\
\verb|owl:intersectionof|            & \XBox         & \XBox          & \XBox       & \XBox                          & \XBox                   & \XBox          & \CheckedBox    \\
\verb|owl:InverseFunctionalProperty| & \CheckedBox   & \XBox          & \XBox       & \CheckedBox                    & \CheckedBox             & \CheckedBox    & \CheckedBox    \\
\verb|owl:inverseOf|                 & \CheckedBox   & \CheckedBox    & \CheckedBox & \CheckedBox                    & \CheckedBox             & \CheckedBox    & \CheckedBox    \\
\verb|owl:IrreflexiveProperty|       & \CheckedBox   & \XBox          & \XBox       & \XBox                          & \XBox                   & \CheckedBox    & \CheckedBox    \\
\verb|owl:maxCardinality|            & \XBox         & \XBox          & \XBox       & \XBox                          & \XBox                   & \XBox          & \CheckedBox    \\
\verb|owl:minCardinality|            & \XBox         & \XBox          & \XBox       & \XBox                          & \XBox                   & \XBox          & \CheckedBox    \\
\verb|owl:ObjectProperty|            & \CheckedBox   & \XBox          & \XBox       & \CheckedBox                    & \XBox                   & \CheckedBox    & \CheckedBox    \\
\verb|owl:oneOf|                     & \XBox         & \XBox          & \XBox       & \XBox                          & \XBox                   & \XBox          & \CheckedBox    \\
\verb|owl:propertyChainAxiom|        & \XBox         & \XBox          & \XBox       & \XBox                          & \XBox                   & \XBox          & \CheckedBox    \\
\verb|owl:propertyDisjointWith|      & \CheckedBox   & \XBox          & \XBox       & \XBox                          & \XBox                   & \CheckedBox    & \CheckedBox    \\
\verb|owl:qualifiedCardinality|      & \XBox         & \XBox          & \XBox       & \XBox                          & \XBox                   & \XBox          & \CheckedBox    \\
\verb|owl:qualifiedMaxCardinality|   & \XBox         & \XBox          & \XBox       & \XBox                          & \XBox                   & \XBox          & \CheckedBox    \\
\verb|owl:qualifiedMinCardinality|   & \XBox         & \XBox          & \XBox       & \XBox                          & \XBox                   & \XBox          & \CheckedBox    \\
\verb|owl:sameAs|                    & \CheckedBox   & \CheckedBox    & \CheckedBox & \CheckedBox                    & \CheckedBox             & \CheckedBox    & \CheckedBox    \\
\verb|owl:someValuesFrom|            & \XBox         & \XBox          & \XBox       & \XBox                          & \CheckedBox             & \XBox          & \CheckedBox    \\
\verb|owl:sourceIndiviual|           & \XBox         & \XBox          & \XBox       & \XBox                          & \XBox                   & \XBox          & \CheckedBox    \\
\verb|owl:SymmetricProperty|         & \CheckedBox   & \XBox          & \CheckedBox & \CheckedBox                    & \CheckedBox             & \CheckedBox    & \CheckedBox    \\
\verb|owl:targetIndividual|          & \XBox         & \XBox          & \XBox       & \XBox                          & \XBox                   & \XBox          & \CheckedBox    \\
\verb|owl:targetValue|               & \XBox         & \XBox          & \XBox       & \XBox                          & \XBox                   & \XBox          & \CheckedBox    \\
\verb|owl:TransitiveProperty|        & \CheckedBox   & \CheckedBox    & \CheckedBox & \CheckedBox                    & \CheckedBox             & \CheckedBox    & \CheckedBox    \\
\verb|owl:unionof|                   & \XBox         & \XBox          & \XBox       & \XBox                          & \XBox                   & \XBox          & \CheckedBox    \\
\verb|rdfs:domain|                   & \CheckedBox   & \CheckedBox    & \CheckedBox & \CheckedBox                    & \CheckedBox             & \CheckedBox    & \CheckedBox    \\
\verb|rdfs:range|                    & \CheckedBox   & \CheckedBox    & \CheckedBox & \CheckedBox                    & \CheckedBox             & \CheckedBox    & \CheckedBox    \\
\verb|rdfs:subClassOf|               & \CheckedBox   & \CheckedBox    & \CheckedBox & \CheckedBox                    & \CheckedBox             & \CheckedBox    & \CheckedBox    \\
\verb|rdfs:subPropertyOf|            & \CheckedBox   & \CheckedBox    & \CheckedBox & \CheckedBox                    & \CheckedBox             & \CheckedBox    & \CheckedBox    \\
\hline
\end{tabular}
\end{table*}

\begin{table}[]
\caption{Vocabulary terms used in LOD snapshot 2015}
\label{tab:fowl}
\centering
\begin{tabular}{|l|c|}
\hline
                                     &  voc terms \\
\hline
\verb|owl:AllDifferent|              & 111       \\
\verb|owl:AllDisjointClasses|        & 21        \\
\verb|owl:AllDisjointProperties|     & 13        \\
\verb|owl:allValuesFrom|             & 126330    \\
\verb|owl:assertionProperty|         & 0         \\
\verb|owl:AsymmetricProperty|        & 0         \\
\verb|owl:cardinality|               & 23910     \\
\verb|owl:complementOf|              & 873       \\
\verb|owl:DatatypeProperty|          & 27471     \\
\verb|owl:differentFrom|             & 784       \\
\verb|owl:disjointUnionOf|           & 0         \\
\verb|owl:disjointWith|              & 3743      \\
\verb|owl:equivalentClass|           & 29708     \\
\verb|owl:equivalentProperty|        & 201       \\
\verb|owl:FunctionalProperty|        & 3730      \\
\verb|owl:hasKey|                    & 5         \\
\verb|owl:hasSelf|                   & 3         \\
\verb|owl:hasValue|                  & 1877      \\
\verb|owl:intersectionOf|            & 2681      \\
\verb|owl:InverseFunctionalProperty| & 94        \\
\verb|owl:inverseOf|                 & 1341      \\
\verb|owl:IrreflexiveProperty|       & 0         \\
\verb|owl:maxCardinality|            & 257371    \\
\verb|owl:minCardinality|            & 455203    \\
\verb|owl:ObjectProperty|            & 40330     \\
\verb|owl:oneOf|                     & 853       \\
\verb|owl:propertyChainAxiom|        & 68        \\
\verb|owl:propertyDisjointWith|      & 4         \\
\verb|owl:qualifiedCardinality|      & 109       \\
\verb|owl:qualifiedMaxCardinality|   & 2         \\
\verb|owl:qualifiedMinCardinality|   & 20        \\
\verb|owl:sameAs|                    & 3967150   \\
\verb|owl:someValuesFrom|            & 4446      \\
\verb|owl:sourceIndiviual|           & 0         \\
\verb|owl:SymmetricProperty|         & 194       \\
\verb|owl:targetIndividual|          & 0         \\
\verb|owl:targetValue|               & 11        \\
\verb|owl:TransitiveProperty|        & 267       \\
\verb|owl:unionOf|                   & 53735     \\
\verb|rdfs:domain|                   & 111865    \\
\verb|rdfs:range|                    & 59252     \\
\verb|rdfs:subClassOf|               & 1339391   \\
\verb|rdfs:subPropertyOf|            & 13416     \\
\hline
\end{tabular}
\end{table}

\begin{table*}[]
\caption{Inference rules for OWL-P properties}
\label{tab:rulesowl1}
\centering
\begin{tabular}{|lll|}
\hline
Conditions &  & Conclusions \\
\hline
\verb|{?S ?P ?O}| & \verb|=>| & \verb|{?S owl:sameAs ?S. ?P owl:sameAs ?P.| \\
                  &           &  \verb| ?O owl:sameAs ?O}.| \\
\verb|{?S owl:sameAs ?O}| & \verb|=>| & \verb|{?O owl:sameAs ?S}.| \\
\verb|{?Q owl:sameAs ?R. ?R owl:sameAs ?P}| & \verb|=>| & \verb|{?Q owl:sameAs ?P}.| \\
\verb|{?S owl:sameAs ?S2. ?S ?P ?O}| & \verb|=>| & \verb|{?S2 ?P ?O}.| \\
\verb|{?P owl:sameAs ?P2. ?S ?P ?O}| & \verb|=>| & \verb|{?S ?P2 ?O}.| \\
\verb|{?O owl:sameAs ?O2. ?S ?P ?O}| & \verb|=>| & \verb|{?S ?P ?O2}.| \\
\verb|{?Q owl:sameAs ?R. ?Q owl:differentFrom ?R}| & \verb|=>| & \verb|{@false}.| \\
\verb|{?P rdf:type owl:FunctionalProperty.| & & \\
\verb| ?Q ?P ?R1. ?Q ?P ?R2}| & \verb|=>| & \verb|{?R1 owl:sameAs ?R2}.| \\
\verb|{?P rdf:type owl:InverseFunctionalProperty.| & & \\
\verb| ?Q1 ?P ?R. ?Q2 ?P ?R}| & \verb|=>| & \verb|{?Q1 owl:sameAs ?Q2}.| \\
\verb|{?P rdf:type owl:IrreflexiveProperty. ?Q ?P ?Q}| & \verb|=>| & \verb|{@false}.| \\
\verb|{?P rdf:type owl:SymmetricProperty. ?Q ?P ?R}| & \verb|=>| & \verb|{?R ?P ?Q}.| \\
\verb|{?P rdf:type owl:AsymmetricProperty. ?Q ?P ?R. ?R ?P ?Q }| & \verb|=>| & \verb|{@false}.| \\
\verb|{?P rdf:type owl:TransitiveProperty. ?Q ?P ?R. ?R ?P ?P}| & \verb|=>| & \verb|{?Q ?P ?P}.| \\
\verb|{?P1 owl:equivalentProperty ?P2. ?Q ?P1 ?R}| & \verb|=>| & \verb|{?Q ?P2 ?R}.| \\
\verb|{?P1 owl:equivalentProperty ?P2. ?Q ?P2 ?R}| & \verb|=>| & \verb|{?Q ?P1 ?R}.| \\
\verb|{?P1 owl:propertyDisjointWith ?P2. ?Q ?P1 ?R. ?Q ?P2 ?R}| & \verb|=>| & \verb|{@false}.| \\
\verb|{?P1 owl:inverseOf ?P2. ?Q ?P1 ?R}| & \verb|=>| & \verb|{?R ?P2 ?Q}.| \\
\verb|{?P1 owl:inverseOf ?P2 . ?Q ?P2 ?R}| & \verb|=>| & \verb|{?R ?P1 ?Q}.| \\
\hline
\end{tabular}
\end{table*}

\begin{table*}[]
\caption{Inference rules for OWL-P classes}
\label{tab:rulesowl2}
\centering
\begin{tabular}{|lll|}
\hline
Conditions &  & Conclusions \\
\hline
\verb|{?A owl:equivalentClass ?B . ?x rdf:type ?A}| & \verb|=>| & \verb|{?x a ?B}.| \\
\verb|{?A owl:equivalentClass ?B . ?x rdf:type ?B}| & \verb|=>| & \verb|{?x a ?A}.| \\
\verb|{?A owl:disjointWith ?B.| & & \\
\verb|{ ?x rdf:type ?A. ?x rdf:type ?B}| & \verb|=>| & \verb|{@false}| \\
\verb|{?C rdf:type owl:Class}| & \verb|=>| & \verb|{?C rdfs:subClassOf ?C. ?C owl:Thing.| \\
                               &           & \verb| ?C owl:equivalentClass ?C.| \\
                               &           & \verb| owl:Nothing rdfs:subClassOf ?C}.| \\
\verb|{?A owl:equivalentClass ?B}| & \verb|=>| & \verb|{?A rdfs:subClassOf ?B. ?B rdfs:subClassOf ?A}.| \\
\verb|{?A rdfs:subClassOf ?B.| & & \\
\verb| ?B rdfs:subClassOf ?A}| & \verb|=>| & \verb|{?A owl:equivalentClass ?B}.| \\
\verb|{?P rdf:type owl:ObjectProperty}| & \verb|=>| & \verb|{?P rdfs:subPropertyOf ?P.| \\
                                        &           & \verb| ?P owl:equivalentProperty ?P}.| \\
\verb|{?P rdf:type owl:DatatypeProperty}| & \verb|=>| & \verb|{?P rdfs:subPropertyOf ?P.| \\
                                          &           & \verb| ?P owl:equivalentProperty ?P}.| \\
\verb|{?P owl:equivalentProperty ?R}| & \verb|=>| & \verb|{?P rdfs:subPropertyOf ?R.| \\
                                        &           & \verb| ?R rdfs:subPropertyOf ?P}.| \\
\verb|{?P rdfs:subPropertyOf ?R.| & & \\
\verb| ?R rdfs:subPropertyOf ?P}| & \verb|=>| & \verb|{?P owl:equivalentProperty ?R}.| \\
\hline
\end{tabular}
\end{table*}

\section{Related Work}
\label{sec:relatedwork}
One of the most important general purpose logic programming language is Prolog \cite{Clocksin2003}. It is declarative, which means
that the program logic is declared in terms of relations, represented as facts and rules. Yet anoder declarative language
is Datalog \cite{Eiter1997}, which is syntactically a subset of Prolog. Apart from the Notation3, there are other rule-based 
inference engines formats for the Semantic Web, such as: FOL-RuleML \cite{Boley2004}, SWRL \cite{Horrocks2004}, RIF \cite{Kifer2008}, 
R-DEVICE \cite{Bassiliades2004}, TRIPLE \cite{Sintek2001}, Jena rule\footnote{\url{http://jena.apache.org/documentation/inference}} and
SPIN \cite{Ryman2014}.

FOL-RuleML (First-order Logic Rule Markup Language) \cite{Boley2004} is a rule language for expressing first-order 
logic for the web. It is a sublanguage of RuleML \cite{Boley2012}. In FOL-RuleML each of rules consists of a set of statements 
called an atom. The atom is a form which consists of objects which are individuals or variables, and a relation between them.

SWRL (Semantic Web Rule Language) \cite{Horrocks2004} is based on OWL \cite{Parsia2012} and Unary/Binary Datalog RuleML, which sublanguage of 
the RuleML. It extends the set of OWL axioms to include Horn-like rules. Logical operators and quantifications supports 
of SWRL are the same as in RuleML. Moreover, RuleML contents can be parts of SWRL content. Axioms may consist of OWL, RDF
and rule axioms. A relation can be an IRI, a data range, an OWL property or a built-in relation. An object can be a variable, 
an individual, a literal value or a blank node.

RIF (Rule Interchange Format) \cite{Kifer2008} is a standard for exchanging rules among disparate systems. It focused on 
exchange rather than developing a single one-fits-all rule language. It can be separated into a number of parts, 
RIF-core \cite{Reynolds2013} which is the common core of all RIF dialects, RIF-BLD (Basic Logic Dialect) \cite{Kifer2013} comprising 
basic dialects (i.e. Horn rules) for writing rules, RIF-PRD \cite{Marie2013} (Production Rule Dialect) for representing production rules and 
RIF-DTB (Datatypes and Built-in Functions) \cite{Polleres2013} comprising a set of datatypes and built-in functions.

R-DEVICE \cite{Bassiliades2004} is a deductive rule language for reasoning about RDF data. In R-DEVICE resources are represented 
as objects and RDF properties are realized as multi-slots. It supports a second-order syntax, where variables can range over 
classes and properties. It provides a RuleML-like syntax.

TRIPLE \cite{Sintek2001} is an RDF rule (query, inference, and transformation) language, with a layered and modular nature. 
It is based on Horn Logic \cite{Horn1951} and F-Logic \cite{Kifer1989}. Rules in TRIPLE are used for transient querying and 
cannot be used for defining and maintaining views.

SPIN (SPARQL Inferencing Notation) \cite{Ryman2014} is a constraint and SPARQL-based rule language for RDF. It can link class with 
queriesto capture constraints and rules which describe the behavior of those classes. SPIN is also a method to represent queries as 
templates. It can represent SPARQL statement as RDF triples. That proposal allows to declare new SPARQL functions.

Jena rule is a rule format used only by inference engine in the Jena framework \cite{Mcbride2002}. The rule language 
syntax is based on RDF. It uses the triple representation, which is similar to Notation3 except that a rule name can be 
specified in a rule. There are not any formula notation, and built-in functions are written in function terms. 

\section{Conclusions}
This paper define how knowledge and logic might be handled on the Semantic Web environment. We present inference rules 
RDF, RDF Schema and OWL. All rules are tested in reasoning engines. Our formalization is based on Notation 3 Logic, which 
extended RDF by logical symbols and created a new Semantic Web logic. Moreover, we propose a lightweight OWL profile called OWL-P.
Our proposed rule will be useful for deductive 
RDF graph stores.

\bibliography{references}

\begin{thebibliography}{10}

\bibitem{Rdfspp2016}
Overview of rdfs++.
\newblock http://franz.com/agraph/support/learning/Overview-of-RDFS++.lhtml.
\newblock Accessed: 2016-01-06.

\bibitem{Arndt2015}
D{\"o}rthe Arndt, Ruben Verborgh, Jos {De Roo}, Hong Sun, Erik Mannens, and Rik
  {Van de Walle}.
\newblock {Semantics of {Notation3} Logic: A solution for implicit
  quantification}.
\newblock In {\em {Proceedings of the 9th International Web Rule Symposium}},
  August 2015.

\bibitem{Bassiliades2004}
Nick Bassiliades and Ioannis Vlahavas.
\newblock {R-device: A deductive rdf rule language}.
\newblock In {\em {Rules and Rule Markup Languages for the Semantic Web}},
  pages 65--80. Springer, 2004.

\bibitem{Clocksin2003}
William Clocksin and Christopher~S Mellish.
\newblock {\em {Programming in PROLOG}}.
\newblock Springer Science \& Business Media, 2003.

\bibitem{Eiter1997}
Thomas Eiter, Georg Gottlob, and Heikki Mannila.
\newblock {Disjunctive datalog}.
\newblock {\em ACM Transactions on Database Systems (TODS)}, 22(3):364--418,
  1997.

\bibitem{Fischer2010}
Florian Fischer, Gulay Unel, Barry Bishop, and Dieter Fensel.
\newblock Towards a scalable, pragmatic knowledge representation language for
  the web.
\newblock In {\em Perspectives of Systems Informatics}, pages 124--134.
  Springer, 2010.

\bibitem{Glimm2011}
Birte Glimm, Aidan Hogan, Markus Krotzsch, and Axel Polleres.
\newblock Owl ld: Entailment ruleset and implementational notes.
\newblock http://semanticweb.org/OWLLD/.

\bibitem{Boley2004}
Mike Dean Benjamin Grosof Michael Sintek Bruce Spencer Said~Tabet {Harold
  Boley} and Gerd Wagner.
\newblock {{FOL RuleML}: The First-Order Logic Web Language}.
\newblock Technical report, November 2004.
\newblock {http://ruleml.org/fol/}.

\bibitem{Boley2012}
Tara Athan Adrian Paschke Adrian Giurca Nick Bassiliades Guido Governatori
  Monica Palmirani Adam Wyner Gen~Zou {Harold Boley} and Zhili Zhao.
\newblock {Specification of Deliberation RuleML 1.01}.
\newblock Technical report, 2012.
\newblock
  {http://wiki.ruleml.org/index.php/Specification\_of\_Deliberation\_RuleML\_1.01}.

\bibitem{Hendler2010}
Jim Hendler.
\newblock Rdfs 3.0.
\newblock In {\em W3C Workshop -- RDF Next Steps}. World Wide Web, 2010.

\bibitem{Horn1951}
Alfred Horn.
\newblock {On sentences which are true of direct unions of algebras}.
\newblock {\em The Journal of Symbolic Logic}, 16(01):14--21, 1951.

\bibitem{Horrocks2004}
Ian Horrocks, Peter~F Patel-Schneider, Harold Boley, Said Tabet, Benjamin
  Grosof, Mike Dean, et~al.
\newblock {SWRL: A semantic web rule language combining OWL and RuleML}.
\newblock {W3C} member submission, World Wide Web Consortium, May 2004.
\newblock {http://www.w3.org/Submission/2004/SUBM-SWRL-20040521/}.

\bibitem{Isele2010}
Robert Isele, J\"{u}rgen Umbrich, Chris Bizer, and Andreas Harth.
\newblock {LDSpider}: An open-source crawling framework for the web of linked
  data.
\newblock In {\em Proceedings of 9th International Semantic Web Conference
  (ISWC 2010) Posters and Demos}, 2010.

\bibitem{Kifer2008}
Michael Kifer.
\newblock {Rule interchange format: The framework}.
\newblock In {\em {Web reasoning and rule systems}}, pages 1--11. Springer,
  2008.

\bibitem{Kifer2013}
Michael Kifer and Harold Boley.
\newblock {{RIF} Basic Logic Dialect (Second Edition)}.
\newblock {W3C} recommendation, World Wide Web Consortium, February 2013.
\newblock {http://www.w3.org/TR/2013/REC-rif-bld-20130205/}.

\bibitem{Kifer1989}
Michael Kifer and Georg Lausen.
\newblock {F-logic: a higher-order language for reasoning about objects,
  inheritance, and scheme}.
\newblock In {\em {ACM SIGMOD Record}}, volume~18, pages 134--146. ACM, 1989.

\bibitem{Ryman2014}
Holger Knublauch, James~A. Hendle, and Kingsley Idehen.
\newblock {{SPIN} - Overview and Motivation}.
\newblock {W3C} member submission, World Wide Web Consortium, February 2011.
\newblock {http://www.w3.org/Submission/2011/SUBM-spin-overview-20110222/}.

\bibitem{Marie2013}
Christian de~Sainte Marie, Adrian Paschke, and Gary Hallmark.
\newblock {{RIF} Production Rule Dialect (Second Edition)}.
\newblock {W3C} recommendation, World Wide Web Consortium, February 2013.
\newblock {http://www.w3.org/TR/2013/REC-rif-prd-20130205/}.

\bibitem{Mcbride2002}
Brian McBride.
\newblock Jena: A semantic web toolkit.
\newblock {\em IEEE Internet computing}, (6):55--59, 2002.

\bibitem{Motik2012}
Boris Motik, Bernardo~Cuenca Grau, Ian Horrocks, Achille Fokoue, and Zhe Wu.
\newblock {OWL} 2 web ontology language profiles (second edition).
\newblock {W3C} recommendation, World Wide Web Consortium, December 2012.
\newblock http://www.w3.org/TR/2012/REC-owl2-profiles-20121211/.

\bibitem{Parsia2012}
Bijan Parsia, Sebastian Rudolph, Markus Kr{\"o}tzsch, Peter Patel-Schneider,
  and Pascal Hitzler.
\newblock {{OWL} 2 Web Ontology Language Primer (Second Edition)}.
\newblock {W3C} recommendation, World Wide Web Consortium, December 2012.
\newblock {http://www.w3.org/TR/2012/REC-owl2-primer-20121211/}.

\bibitem{Polleres2013}
Axel Polleres, Michael Kifer, and Harold Boley.
\newblock {{RIF} Datatypes and Built-Ins 1.0 (Second Edition)}.
\newblock {W3C} recommendation, World Wide Web Consortium, February 2013.
\newblock {http://www.w3.org/TR/2013/REC-rif-dtb-20130205/}.

\bibitem{Reynolds2013}
Dave Reynolds, Michael Kifer, Axel Polleres, Harold Boley, Adrian Paschke, and
  Gary Hallmark.
\newblock {{RIF} Core Dialect (Second Edition)}.
\newblock {W3C} recommendation, World Wide Web Consortium, February 2013.
\newblock {http://www.w3.org/TR/2013/REC-rif-core-20130205/}.

\bibitem{Schmachtenberg2014}
Max Schmachtenberg, Christian Bizer, and Heiko Paulheim.
\newblock {Adoption of the linked data best practices in different topical
  domains}.
\newblock In {\em {The Semantic Web--ISWC 2014}}, pages 245--260. Springer,
  2014.

\bibitem{Sintek2001}
Michael Sintek and Stefan Decker.
\newblock {TRIPLE-An RDF Query, Inference, and Transformation Language.}
\newblock In {\em {INAP}}, pages 47--56, 2001.

\bibitem{Sperberg-McQueen2012}
Michael Sperberg-McQueen, Ashok Malhotra, Paul~V. Biron, Sandy Gao, Henry
  Thompson, and David Peterson.
\newblock {{W3C} XML Schema Definition Language ({XSD}) {1.1} Part 2:
  Datatypes}.
\newblock {W3C} recommendation, World Wide Web Consortium, April 2012.
\newblock {http://www.w3.org/TR/2012/REC-xmlschema11-2-20120405/}.

\bibitem{Terhorst2005}
Herman~J ter Horst.
\newblock Completeness, decidability and complexity of entailment for rdf
  schema and a semantic extension involving the owl vocabulary.
\newblock {\em Web Semantics: Science, Services and Agents on the World Wide
  Web}, 3(2):79--115, 2005.

\bibitem{Wood2014}
David Wood, Markus Lanthaler, and Richard Cyganiak.
\newblock {{RDF} 1.1 Concepts and Abstract Syntax}.
\newblock {W3C} recommendation, World Wide Web Consortium, February 2014.
\newblock {http://www.w3.org/TR/2014/REC-rdf11-concepts-20140225/}.

\end{thebibliography}
\bibliographystyle{plain}

\begin{IEEEbiography}[{\includegraphics[width=1in,height=1.25in,clip,keepaspectratio]{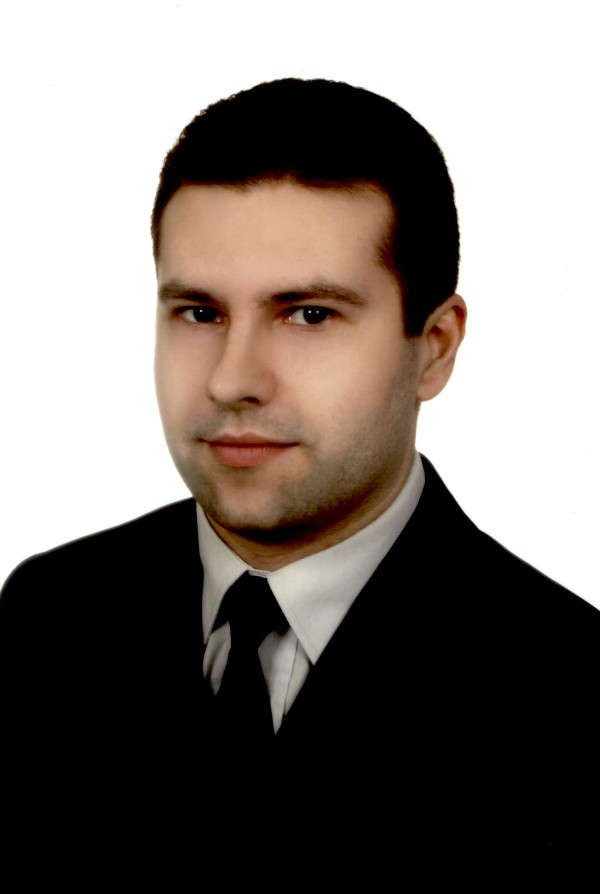}}]{Dominik Tomaszuk}
Dr. Dominik Tomaszuk is a researcher at the University of Bialystok. His current research focuses on RDF, data streams, and NoSQL databases.
\end{IEEEbiography}
\end{document}